\documentclass[sigconf]{acmart}

\usepackage{booktabs} 
\usepackage{subfigure}

\copyrightyear{2019}
\acmYear{2019}
\setcopyright{iw3c2w3}
\acmConference[WWW '19]{Proceedings of the 2019 World Wide Web Conference}{May 13--17, 2019}{San Francisco, CA, USA}
\acmBooktitle{Proceedings of the 2019 World Wide Web Conference (WWW '19), May 13--17, 2019, San Francisco, CA, USA}
\acmPrice{}
\acmDOI{10.1145/3308558.3313444}
\acmISBN{978-1-4503-6674-8/19/05}

\usepackage{balance}

\begin{document}

\title{Dressing as a Whole: Outfit Compatibility Learning \\
Based on Node-wise Graph Neural Networks}
\author{Zeyu Cui}
\affiliation{Institute of Automation, Chinese Academy of Sciences}
\affiliation{University of Chinese Academy of Sciences}
\email{zeyu.cui@nlpr.ia.ac.cn}
\author{Zekun Li}
\affiliation{Institute of Information Engineering, Chinese Academy of Sciences}
\affiliation{University of Chinese Academy of Sciences}
\email{lizekunlee@gmail.com}
\author{Shu Wu}
\affiliation{Institute of Automation, Chinese Academy of Sciences}
\affiliation{University of Chinese Academy of Sciences}
\email{shu.wu@nlpr.ia.ac.cn}
\author{Xiaoyu Zhang}
\affiliation{Institute of Information Engineering, Chinese Academy of Sciences}
\email{zhangxiaoyu@iie.ac.cn}
\author{Liang Wang}
\affiliation{Institute of Automation, Chinese Academy of Sciences}
\affiliation{University of Chinese Academy of Sciences}
\email{wangliang@nlpr.ia.ac.cn}
\thanks{The first two authors Zeyu Cui and Zekun Li contribute to this work equally and are listed as joint first authors.
Shu Wu and Xiaoyu Zhang are both corresponding authors.}



\begin{abstract}
With the rapid development of fashion market, the customers' demands of customers for fashion recommendation are rising. In this paper, we aim to investigate a practical problem of fashion recommendation by answering the question
``which item should we select to match with the given fashion items and form a compatible outfit".
The key to this problem is to estimate the outfit compatibility.
 Previous works which focus on the compatibility of two items or represent an outfit as a sequence fail to make full use of the complex relations among items in an outfit.
 To remedy this, we propose to represent an outfit as a graph. In particular, we construct a \emph{Fashion Graph}, where each node represents a category
 and each edge represents interaction between two categories.
 Accordingly, each outfit can be represented as a subgraph by putting items into their corresponding category nodes.
 To infer the outfit compatibility from such a graph, we propose \emph{Node-wise Graph Neural Networks} (NGNN) which can better model node interactions and learn better node representations.
  In NGNN, the node interaction on each edge is different, which is determined by parameters correlated to the two connected nodes.
 An attention mechanism is utilized to calculate the outfit compatibility score with learned node representations.
NGNN can not only be used to model outfit compatibility from visual or textual modality but also from multiple modalities.
 We conduct experiments on two tasks: (1) Fill-in-the-blank:  suggesting an item that matches with existing components of outfit;
 (2) Compatibility prediction: predicting the compatibility scores of given outfits.
 Experimental results demonstrate the great superiority of our proposed method over others.

\end{abstract}

%
%



\maketitle

\section{Introduction} \label{sect:intro}
Clothing plays an increasingly significant role in human's social life, as a proper outfit can enhance one's beauty and display the personality.
However, not everyone has a strong fashion sensitivity.
 ``How to make a suitable and beautiful outfit?'' has become a daily headache for many people.  Solving this problem requires modeling human notions of the compatibility. Here we aim to propose an effective method to predict the compatibility of outfits and thus help people make proper outfits.

\begin{figure}[t]
\centering
\includegraphics[width=1\linewidth]{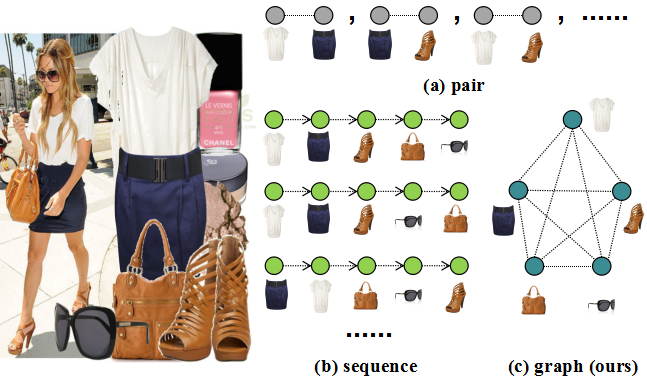}
\caption{
Different representations of an outfit. Left is the outfit. (a), (b) and (c) in the right are its pair, sequence and graph representations respectively. Circles represent the items and links represent the relations.}
\label{fig:intro}
\end{figure}

\begin{figure*}[t]
\centering
\includegraphics[width=1\linewidth]{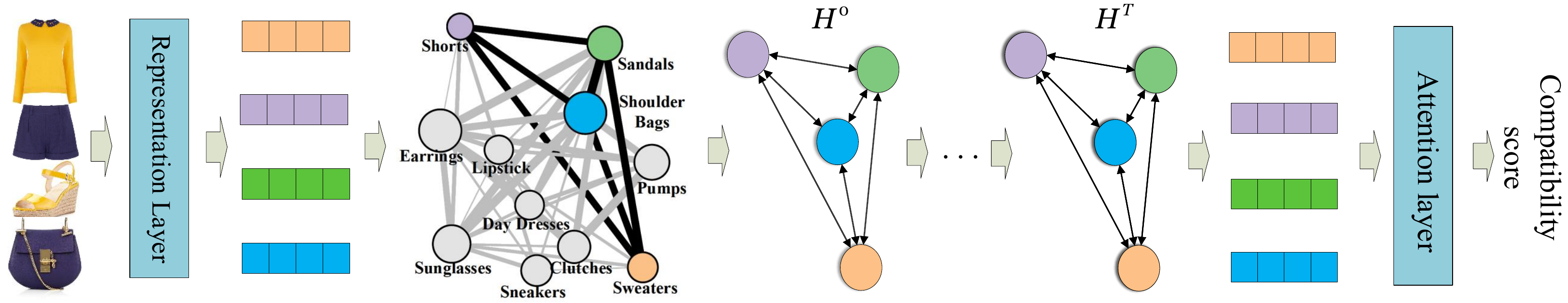}
\caption{
Overview of the proposed method. Based on the dataset, we first construct a \emph{Fashion Graph}, where each node represents a category and each edge represents the interaction between two nodes. An outfit (e.g. sweater, short, sandals, shoulder bag) can be represented as a subgraph. We then design NGNN to model the node interactions and learn node representations. An attention layer is finally utilized to calculate the compatibility score. }
\label{fig:framework}
\end{figure*}

Previous works on compatibility learning mainly focused on the compatibility of two items. These works followed the idea to map the items into a style space and estimate the distance between style vectors of items.
 Veit \emph{et.al} used SiameseCNNs \cite{veit2015learning,Bell2015Learning} to transform items from the image space to the style space.
 Mcauley \emph{et.al} \cite{mcauley2015image} proposed to use Low-rank Mahalanobis Transformation to map items into the style space. Lately, some works proposed to model the compatibility of items by mapping items into several style spaces \cite{he2016learning,shih2018compatibility}.
Recently, there appear a few works directly measure the compatibility of outfit.
Li \emph{et.al} \cite{li2017mining} proposed to represent an outfit as a sequence and adopt Recurrent Neural Network (RNN) \cite{mikolov2010recurrent,mikolov2011extensions,hochreiter1997long,gers1999learning,sundermeyer2012lstm,liu2016predicting,cui2018mv} to model the compatibility.
Similarly, another work \cite{han2017learning} further represented an outfit as a bidirectional sequence with a specific order (i.e., top to bottom and then to accessories), and used Bidirectional Long Short Term Memories (Bi-LSTMs) to predict the next item in an outfit.

Solving the task of modeling outfit compatibility relies on an appropriate outfit representation. The previous works take advantage of two kinds of outfit representations: \emph{pair representation} \cite{veit2015learning,mcauley2015image,he2016learning,shih2018compatibility}
and \emph{sequence representation} \cite{han2017learning}.
We show different outfit representations in Figure \ref{fig:intro}.
From (a) we can see that representing the outfit as pairs separately cannot reflect the complex relations among multiple items.
 In term of \emph{sequence representation} in (b), there doesn't exist a fixed order of items. More importantly, the relations among items in an outfit are not ordered since an item has relations with not only its prior or rear item in the sequence.
 To remedy this, we propose a new outfit representation: \emph{graph representation}. As shown in (c), it can better reflect the dense and complex relations among multiple items in an outfit.

In particular, we construct a $\emph{Fashion Graph}$ based on Polyvore dataset, where each node represents a category and each edge represents the interaction between two nodes. If two categories have matching relation (i.e., they appear in the same outfit), there should be interactions between them. We have two directed edges between them in $\emph{Fashion Graph}$ since the interactions in two directions ought to be different. For example, one may select a pair of socks to match the shoes but not select shoes for socks in particular.
A compatible outfit should have a key property: the categories of items in it are not overlapped (e.g., there won't be two pairs of shoes in an outfit).
Thus, by putting each item into its corresponding node, an outfit can be represented as a subgraph of \emph{Fashion Graph}.
The task of modeling outfit compatibility can be accordingly converted to a graph inference problem.

A possible method is using Graph Neural Networks (GNN) to model the node interactions and infer compatibility from the graph.
GNN is an approach to applying neural networks to graph-structured data, first proposed by Gori \emph{et al.} \cite{gori2005new}.
It can be applied on most kinds of graphs, including directed, undirected and cyclic graphs. However, it has trouble in propagating information across a long range in a graph. To address this, Li \emph{et al.} \cite{li2015gated} proposed Gated Graph Neural Network (GGNN) by introducing a Gated Recurrent Units (GRU) for updating. Nevertheless, in GGNN, the nodes interact with others by communicating their state information on the edges in a fixed way, which leads to a limitation that it has difficulty in modeling flexible and complex node interactions.

To address this limitation, we propose a novel model NGNN to better model node interactions and better infer the compatibility from graph.
In NGNN, the communication of state information on each edge is different.
Specifically, when two nodes communicate state information on the edge, the state information will first be transformed by a transformation function determined by parameters correlated to the two nodes.
In this way, NGNN can model edge-wise interactions with node-wise parameters.
Since the parameters are correlated to nodes instead of numerous edges, it can greatly reduce the parameter space.
During training process, NGNN only deals with the nodes in the subgraph each time, which reduces the time complexity as well. Moreover, when there are more data to train the model, we only need to update the parameters of these concerned nodes without changing the edges or other nodes, which is effective in real applications.
With the learned node representations, we utilize an attention mechanism to calculate a graph-level output to serve as the outfit compatibility score.


The framework of our proposed method is shown in Figure \ref{fig:framework}.
We first construct a \emph{Fashion Graph} where an outfit can be represented as a subgraph. Then we use the proposed novel model NGNN to model node interactions and learn the node representations.
Finally, we predict the compatibility score by calculating a graph-level output using an attention layer.
This proposed method can model outfit compatibility from visual, textual or any single modality with a channel of input features.
By using two channels of input to NGNN, it can also be used to jointly model the outfit compatibility from visual and textual modality.

We conduct experiments on two tasks: (1) Fill-in-the-blank: suggesting an item that matches with existing components of outfit;
 (2) Compatibility prediction: predicting the compatibility scores of given outfits.
 Experiment results on the Polyvore dataset demonstrate the great superiority of our method over others.
 The code and dataset of our work have been released\footnote{\url{https://github.com/CRIPAC-DIG/NGNN}}.

Our main contributions can be summarized in threefold:
\begin{itemize}
\item To the best of our knowledge, this is the first attempt to represent an outfit as a graph, which can better capture the complex relations among multiple items in an outfit.
\item We propose a novel model NGNN which can better model node interactions in the graph and learn better node representations.
	Taking advantage of NGNN, we can not only model outfit compatibility from visual or textual modality, but also from multiple modalities.
\item Experimental results on a real-world dataset (Polyvore dataset) demonstrate the great superiority of our method over others.
\end{itemize}

\section{Related Work}

\subsection{Fashion Compatibility Modeling}
There have been many studies focusing on modeling the compatibility of clothing recently, since it is the key to fashion recommendation.
McAuley \emph{et al.} assumed that there existed a latent style space where compatible items stayed close. They used Low-rank Mahalanobis Transformation (LMT) to map items into a latent space and measured the distance there \cite{mcauley2015image}. Following that, Veit \emph{et al.} proposed to learn the distance metric with an end-to-end SiameseCNNs \cite{veit2015learning}. Some other works mapped items into several latent spaces and jointly modeled the distances in these latent spaces, which can measure the compatibility of items in different aspects \cite{he2016learning,shih2018compatibility,chen2018Dress}.
However, these works only focus on the compatibility of two items instead of a whole outfit.

Recently, some studies have been trying to model the compatibility of a whole outfit directly.
Li \emph{et~al.} extracted multi-modal information from items and evaluated the outfit compatibility with an Recurrent Neural Network (RNN) \cite{li2017mining}. Then they constructed an outfit recommendation by selecting the items which achieve the highest compatibility scores along with the given items.
Han \emph{et~al.} represented an outfit as a sequence with a specific order. Then they used bidirectional LSTMs to predict the next item given a set of items and also the compatibility scores of outfits \cite{han2017learning}.
Modeling outfit compatibility relies on a proper representation of the outfit, which can reflect the complex relations among multiple items. In this work, we represent an outfit as a graph, which is proved to be more effective than pairs and sequence in the later experiments.

Categorical information plays a dominant role in the representation of an item.
In the above conventional methods modeling items in a common visual latent space, the items of same categories tends to be close, which makes it harder to reflect the style information.
For example, in the common visual latent space, the similarity between suit pants and leather shoes is much smaller than the similarity between suit pants and jeans.
In Sherlock \cite{he2016sherlock}, the embedding matrices for transferring visual features to style features vary among different categories.
Liu \emph{et al.}  proposed that items consist of two components: style and category \cite{liu2017deepstyle}.
In this work, we utilize the categorical information directly by putting items into its corresponding category nodes in the \emph{Fashion Graph}.

\subsection{Graph Neural Networks}
There have been various ways to deal with graph-structured data. Some works convert graph-structured data into sequence-structured data to deal with. Inspired by word2vec \cite{mikolov2013distributed},
Perozzi \emph{et~al.} proposed an unsupervised DeepWalk algorithm to learn node embedding in graph based on random walks \cite{perozzi2014deepwalk}. After that, Tang \emph{et~al.} proposed a network embedding algorithm LINE \cite{tang2015line}, followed by node2vec \cite{Grover2016node2vec}.
Some works apply neural networks on the graph-structured data directly. Duvenaud \emph{et~al.} designed a special hashing function so that Convolutional Neural Network (CNN) can be used on the graphs \cite{duvenaud2015convolutional}.
Kipf \emph{et al.} proposed a semi-supervised method on graphs based on an efficient variant of CNN \cite{Kipf2016Semi}.

Gori \emph{et al.} first proposed Graph Neural Networks (GNN) \cite{gori2005new}, another approach to applying neural networks directly to graph-structured data.
GNN can be applied on most kinds of graphs, including directed, undirected and cyclic graphs.
Feed-forward neural networks are applied to every node in the graph recurrently.
The nodes interact with others by communicating their state information.
Thus, the hidden state of nodes will update dynamically.
Scarselli \emph{et al.} utilized multi-layer perceptions (MLP) to update the hidden state of nodes \cite{scarselli2009graph}.
However, there exists problem in propagating information across a long range in a graph.
To remedy this, Li \emph{et al.} proposed Gated Graph Neural Network (GGNN) by introducing a Gated Recurrent Units (GRU) for updating \cite{li2015gated}.
GGNN has been utilized for various tasks, such as image classification \cite{marino2017more}, situation recognition \cite{li2017situation}, recommendation \cite{Wu2018Session} and script event prediction \cite{Zhongyang2018Constructing}.
Nevertheless, it has difficulty in modeling complex node interactions since the communication of state information on the edges are in a fixed way.
To address this limitation, we propose a novel model NGNN, which can better model node interactions so as to better infer outfit compatibility from the graph.
In NGNN, the communication of state information on each edge is different. It can model edge-wise interactions with node-wise parameters.

\section{Dataset and Features}
In this section, we introduce the dataset we use and how we extract the visual and textual features of items, which are used as input of our proposed method.

\subsection{Polyvore Dataset} \label{sect:dataset}
\begin{figure}[t]
  \centering
  \includegraphics[scale=0.39]{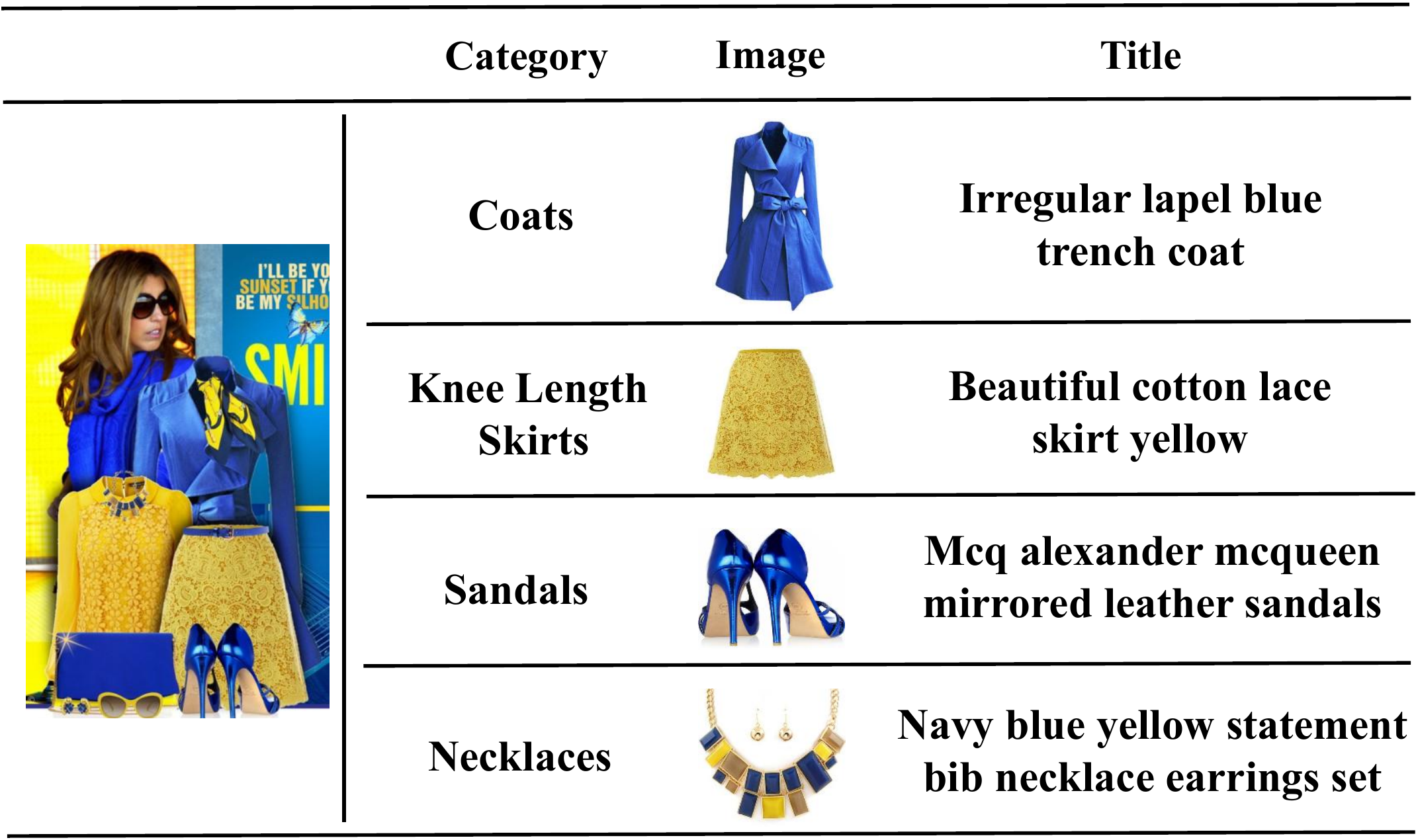}
  \caption{Examples of outfit composition on Polyvore.com. Each item has a clear image, titles and category information.}\label{fig:outfit_example}
\end{figure}

Polyvore dataset released by \cite{han2017learning} was collected from a popular fashion website Polyvore.com\footnote{\url{http://www.polyvore.com/}.}, where fashion stylists can share the outfits they created to the public. As shown in Figure \ref{fig:outfit_example}, items in the outfits have rich information including clear images, category information, titles, numbers of likes, etc.
We only use the images, titles and category information of items in this paper.
The dataset has been utilized by several works on fashion analysis \cite{Hu2015Collaborative,
li2017mining,song2017neurostylist}.

There are totally 21889 outfits covering 380 categories in the Polyvore dataset, which was split into three non-overlapping sets, 17316 for training, 1497 for validation and 3076 for testing.
Following \cite{song2017neurostylist},
there are many categories appearing too few times in the dataset such as \emph{Toys}, \emph{Furniture} and \emph{Cleaning}, which are not actually fashion categories. To ensure the training quality, we keep categories which appear more than 100 times in the dataset and 120 categories remain.
To ensure the integrity of outfits, we filtered out the outfits consisting of less than $3$ items.
Finally, there remain $16983$ outfits in the training set, $1497$ in the validation set and $2697$ in the test set, with totally $126054$ items covering $120$ categories. The maximum size of outfits is $8$ and the average size is $6.2$.

\subsection{Features} \label{sect:feature}
We extract the visual features using the images of items and textual features using titles. We will model the outfit compatibility using these features of items as described in the following.

\subsubsection{Visual Feature}
We utilize the advanced deep convolutional neural network GoogleNet InceptionV3 model~\cite{szegedy2016rethinking} to extract the visual features, which has been proved to be effective in image representation learning \cite{han2017learning,lee2017style2vec,sharif2014cnn,donahue2014decaf}.
In particular, we feed the images of items into the model and adopted the linear layer output as the visual feature. The visual feature of each item is a 2048-dimensional vector.

\subsubsection{Textual Feature}
Following \cite{song2017neurostylist}, we first construct a vocabulary based on the words of the titles in the dataset.
 Due to the noise of these user-generated titles, we first filter out the words which appear in less than 5 items. Since there are many meaningless words such as `a', `an' and `de', we then filter out words with less than 3 characters.
 Finally we obtain a vocabulary with 2757 words and therefore we represented the textual feature of each item as a 2757-dimensional Boolean vector.

\section{Proposed Method}
We first formulate the problem. Then we introduce how to construct the \emph{Fashion Graph} and model outfit compatibility from the graph with NGNN. We also introduce how to jointly model outfit compatibility from multiple modalities with NGNN. Finally, we describe the training strategy and analyze our model.

\subsection{Problem Formulation}
In this paper, we aim to model the compatibility of outfits.
We have a set of outfits $\mathcal{S} = \left \{s_{1}, s_{2}, s_{3}, ... \right \}$ in the training set. The total item set is $\mathcal{V}$.
Given an outfit $s$ consisting of $\left | s \right |$ items (each item has an image and textual description), we aim to predict the compatibility score $x_{s}$ of the outfit. The score can be used for various tasks of fashion recommendation.

We introduce some notations which will be used in the following.
For each item $v_{i} \in \mathcal{V}$, its feature is $f_{i}$ and the  representation of $f_{i}$ in the latent space is $r_{i}$. The category of $v_{i}$ is $c_{i}$. The corresponding node of $c_{i}$ in the \emph{Fashion Graph} is $n_{i}$, while the state of node $n_{i}$ in NGNN is $h_{i}$.

\subsection{Graph Construction} \label{sect:gi}
\begin{figure}[t]
\centering
\includegraphics[width=0.85\linewidth]{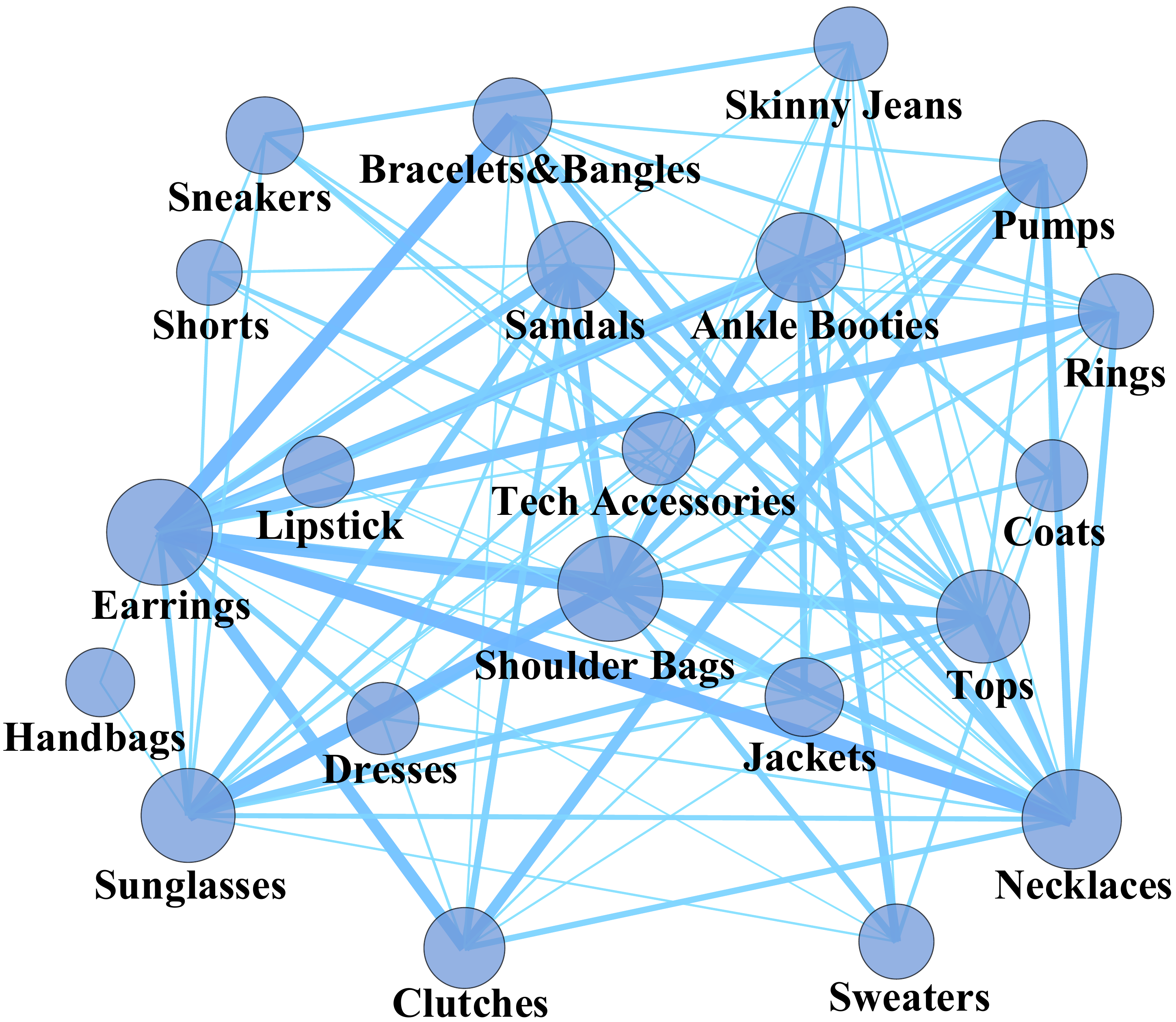}
\caption{
Illustration of the most popular categories in Polyvore dataset. Each circle denotes a category, and each link denotes the matching relation between two categories (i.e., they have appeared in the same outfit).
The areas of circles and the widths of the links are proportional to the number of fashion items in the corresponding categories and the co-occurrence frequency between categories.
}
\label{fig:dataset_category}
\end{figure}

In Figure \ref{fig:dataset_category}, we illustrate the most popular categories in the Polyvore dataset. Each circle denotes a category, and each link denotes the matching relation between two categories (i.e., the two categories have appeared in the same outfit).
The areas of circles and the widths of the links are proportional to the number of fashion items in the corresponding categories and the co-occurrence frequency between categories.

Similar to Figure \ref{fig:dataset_category}, we construct a \emph{Fashion Graph} $\mathcal{G} = (\mathcal{N}, \mathcal{E})$ according to the training dataset.
Each node in the graph $n_{i} \in \mathcal{N}$ represents a category $c_{i} \in \mathcal{C}$, so that $\left | \mathcal{N} \right | = \left | \mathcal{C} \right |$. The edge represents interaction between nodes.
If two categories have matching relation in the training dataset, there are two directed edges in reverse directions between the two corresponding nodes in \emph{Fashion Graph}, since the interactions in two directions ought to different. For example, one may select a pair of socks to match the shoes but less likely to select a pair of shoes to match socks particularly.

According to the training dataset, we calculate the weights of directed edges as follow,
 \begin{equation} \label{uniform_A}
w(n_{i}, n_{j}) = \frac{Count_{c_{i}, c_{j}}/Count_{c_{j}}}{\sum_{k} Count_{c_{i}, c_{k}}/Count_{c_{k}}},
\end{equation}
where $w(n_{i}, n_{j})$ denotes the weight of edge from $n_{i}$ to $n_{j}$, $Count_{c_i, c_j}$ is the co-occurrence frequency of category $c_{i}$ and $c_{j}$, 
and $Count_{c_{j}}$ is the occurrence frequency of category $c_{j}$.

The constructed \emph{Fashion Graph} has 120 nodes and 12500 directed and weighted edges.
By putting each item into its corresponding node, an outfit can be represented as a subgraph of \emph{Fashion Graph}. Then we use NGNN to infer the compatibility from the subgraph as described in the next section.

\subsection{Node-wise Graph Neural Network}
%

Our proposed method NGNN consists of three steps. The first step is to learn the initial node state. The second step is to model node interactions and update node state. The third step is to calculate the compatibility score with attention mechanism.

\subsubsection{Learning Initial Node State}\label{sect:input}
In NGNN, each node $n_{i}$ is associated with a hidden state vector $h_{i} $, which will be updated dynamically.
The input of NGNN is features (either visual or textual features) of items, which can be used to initialize the state of their corresponding nodes.
For each item $v_{i} \in s$, we first map its feature $f_{i}$ to a latent style space with the size of $d$.
Considering the difference among the categories, we have a linear mapping matrix $W_{h}^{i}$ for each category $c_{i}$. Thus, we can obtain the representation of item $v_{i}$ in the latent space as follows,
\begin{equation} \label{latent}
r_{i} = W_{h}^{i}f_i.
\end{equation}
 We then use the representation to initialize the hidden state
$h_{i}^{0}$  of its category corresponding node $n_{i}$ as,
\begin{equation} \label{initial1}
h_{i}^{0} = tanh(r_{i}).
\end{equation}
Thus, the size of node state is also $d$.

\subsubsection{Modeling Node Interactions}

Each time the nodes communicate (i.e., propagate their state information to others and receive others' information), which is called a propagation step, the state of nodes will be updated.

In the traditional GGNN, at propagation step $t$, the nodes will receive sum of neighbors' state as,
 \begin{equation} \label{ggnn_original}
a_{i}^{t} = \sum_{n_{j} \rightarrow n_{i} \in \mathcal{E}} \mathbf{A}[n_{j}, n_{i}] W_{p} h_{j}^{t-1} + b_{p},~
\end{equation}
where $W_{p}$ and $b_{p}$ are weights and biases of a shared linear transformation on all the edges. $\mathbf{A}$ is the adjacency matrix:
\begin{equation} \label{a}
\mathbf{A}[n_{i}, n_{j}]=\begin{cases}
 & w(n_{i}, n_{j}),  \text{ if }  n_{i}\rightarrow n_{j} \in \mathcal{E}, \\
 & 0, \text{ others }.
\end{cases}
\end{equation}
 The traditional GGNN use the same $W_{p}$ and $b_{p}$ to model interactions between different nodes.
 However, it fails to model the complex and flexible interactions between them.
Since the node interaction on each edge ought to be different, we aim to model edge-wise interaction. There is a straight-forward way to achieve edge-wise node interactions, that is to let each edge has its own transformation weight and bias $W_{p}$ and $b_{p}$. However, it will cost a great parameter space with numerous edges in a graph and cannot be applied on large scale graphs.

 To remedy the problem above, we have an output matrix $W_{out}^{i}$ and an input matrix $W_{in}^{i}$ for each node $n_{i}$ in NGNN. The transformation function of edge $n_{i} \rightarrow n_{j}$ from node $n_{i}$ to node $n_{j}$ could be written as
\begin{equation} \label{W_inout}
W_{p}^{n_{i} \rightarrow n_{j}} = W_{out}^{i}W_{in}^{j}.
\end{equation}
Thus, the Equation \ref{ggnn_original} could be rewritten as, 
\begin{equation} \label{ggnn_1}
a_{i}^{t} = \sum_{n_{j} \rightarrow n_{i} \in \mathcal{E}} \textbf{A}[n_{j}, n_{i}]W_{out}^{j}W_{in}^{i} h_{j}^{t-1} + b_{p}.
\end{equation}
After receiving state information $a_{i}^{t}$, the state of node $n_{i}$ will be updated as following,

\begin{align}
& z_{i}^{t} =  \sigma(W_{z} a_{i}^{t} + U_{z}h_{i}^{t-1} + b_{z}),\\
& r_{i}^{t} =  \sigma(W_{r} a_{i}^{t} + U_{r}h_{i}^{t-1} + b_{r}),\\
& \tilde{h}_{i}^{t} =  tanh(W_{h} a_{i}^{t} + U_{h}(r_{i}^{t} \odot h_{i}^{t-1}) + b_{h}), \\
& h_{i}^{t} = \tilde{h}_{i}^{t} \odot z_{i}^{t} + h_{i}^{t-1} \odot (1-z_{i}^{t}),~
\end{align}
where, $W_{z}$, $W_{r}$, $W_{h}$, $b_{z}$, $b_{r}$, $b_{h}$ are weights and biases of the updating function, which are similar to Gated Recurrent Unit (GRU) \cite{li2015gated}. $z_{i}^{t}$ and $r_{i}^{t}$ are update gate vector and reset gate vector, respectively.

After $T$ propagation steps, we can obtain the final state of nodes, which are also the final node representations.
 Then we can generate a graph-level output to serve as the compatibility score $x_{s}$ of outfit $s$. We adopt the attention mechanism described in next section to compute the graph-level output.

\subsubsection{Compatibility Calculating with Attention Mechanism}
We contend two viewpoints on how fashion item influence the outfit compatibility here.
The first viewpoint is that items have different influence on an outfit. For example, an improper top and an improper necklace have different influence to the outfit influence.
The second is that the same item plays different roles in different outfits. For example, a coat maybe suitable for an outfit and promote the compatibility, while it will decrease the compatibility together with summer clothes.
Thus, we aim to design an attention mechanism to model the influence of items on the outfit compatibility.
Since a node receives state information from other nodes, their node representations are aware of the global information (i.e., other nodes' state). Similar with \cite{li2015gated}, we utilize self-attention  \cite{Tan2017Deep,vaswani2017attention} to calculate the graph-level output and also the outfit compatibility score as following,
\begin{equation} \label{self_attention}
x_{s} = \sum_{i=1}^{\left | s \right |} \sigma(\theta(h_{i}^{t})) \cdot \alpha(\delta(h_{i}^{t})).
\end{equation}
$\theta(\cdot)$ and $\delta(\cdot)$ are two perception networks to output a real value.
$\theta(\cdot)$ is used to model the weights of items (i.e., importance of items' influence on the outfit compatibility), and $\delta(\cdot)$ is used to model the compatibility score of items 
(i.e., how the item goes with the other items in the outfit).
$\alpha (\cdot)$ and $\sigma(\cdot)$ are leaky relu and sigmoid activate functions $\alpha(x)=max(0.01x, x), \sigma(x)=1/(1+e^{-x})$.

\subsection{NGNN with Multi-modal Input} \label{sect:multi}
As described above, with the input of visual features or textual features of items, NGNN can be used to model the outfit compatibility from visual or textual modality. We here introduce how to jointly model the outfit compatibility from both visual and textual modality with NGNN.

 We use two channels of NGNN, the input of a channel is visual features, and the input of another channel is textual features.
 Specifically, for each outfit $s$, we input the visual features and textual features to the two channels of NGNN respectively and obtain a visual compatibility score $x_s^{vis}$ and a textual compatibility score $x_{s}^{txt}$. We calculate the final outfit compatibility score $x_{s}$ as follow,
\begin{equation} \label{compatibility}
x_{s}=\beta x_{s}^{vis}+(1-\beta)x_{s}^{txt}
\end{equation}
where $\beta$ is a non-negative trade-off parameter.

As described in Section \ref{sect:input}, for each item $v_{i}$, its visual feature $f_{i}^{vis}$ and textual feature $f_{i}^{txt}$ are mapped into a shared latent space respectively. We denote their representations in latent space as $r_{i}^{vis}$ and $r_{i}^{txt}$. Considering the coherent relation between the visual and textual information of items, we use a regularization to ensure the consistency between the visual and textual feature of the same item in the shared latent space as follow,
\begin{equation} \label{consistency}
\mathcal{L}_{con} = \sum_{i=1}^{\left | s \right |} \left \| r_{i}^{vis} - r_{i}^{txt} \right \|^{2}
\end{equation}

\subsection{Training Strategy}

We denote outfits in the training dataset as positive outfits.
For each outfit $s = \left \{v_{1}, v_{2},...,v_{p},...\right \} \in \mathcal{S}$ , we randomly select an item $v_{p}$ and replace it with a random item $v_{q} \in \mathcal{V}$ to form a negative outfit $s^{-}=\left \{v_{1}, v_{2},...,v_{q},... \right \}$. Thus, we can obtain a dataset $\mathcal{P}_{train}$ consisting of these pairs ($s, s^{-}$).
Similar to Bayesian Personalized Ranking \cite{rendle2009bpr}, we assume that the positive outfit has higher compatibility than the negative one. Therefore, we have the following objective function,
\begin{equation} \label{bpr_func}
\mathcal{L}_{bpr} = \sum_{(s,s^{-}) \in \mathcal{P}_{train}} -\ln \sigma(x_{s} - x_{s^{-}}).
\end{equation}
The objective function of NGNN is:
\begin{equation} \label{obj_func}
\mathcal{L}_{1} = \mathcal{L}_{bpr} + \frac{\lambda}{2}\left \| \Theta \right \|^{2},
\end{equation}
where $\Theta$ refers to the set of parameters and
$\lambda$ is the hyper-parameter of L2 regularizer, which is designed to avoid overfitting.

When it comes to NGNN with multi-modal input, we have an additional loss function $\mathcal{L}_{con}$ as described in Section \ref{sect:multi}. Thus, we have the following objective function,
\begin{equation} \label{obj_func}
\mathcal{L}_{2} = \mathcal{L}_{bpr} + \mathcal{L}_{con} + \frac{\lambda}{2}\left \| \Theta \right \|^{2}.
\end{equation}

We only train NGNN on the subgraph of outfit each time. That is to say, only the nodes in the subgraph are trained and only their parameters are updated each time, which greatly reduces time complexity.
When we have more data to train the model, we only need to update the parameters of concerned nodes without changing the parameters of edges or other nodes, which is effective in real applications.

\begin{figure}[htbp]
\centering

\subfigure[Parameter Size]{
\begin{minipage}[b]{0.24\textwidth}
\label{fig:parameter_space} 
\includegraphics[width=1\textwidth]{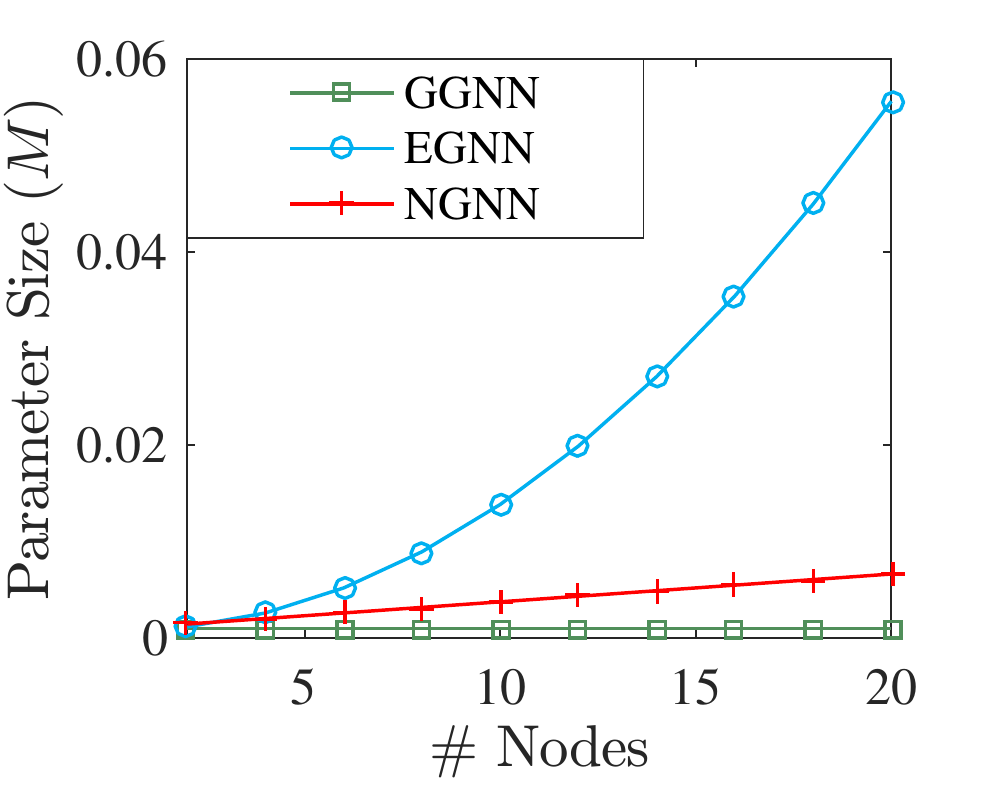}

\end{minipage}%
}%
\subfigure[Running Time]{
\begin{minipage}[b]{0.24\textwidth}
\label{fig:time_consuming} 
\includegraphics[width=1\textwidth]{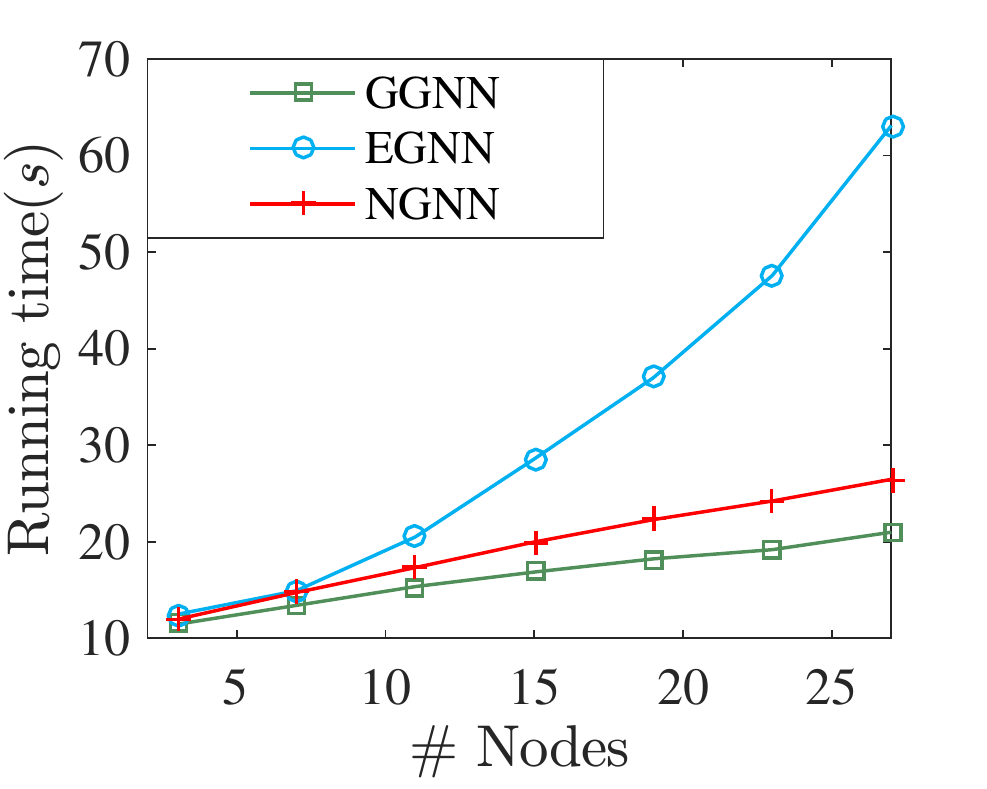}
\end{minipage}%
}%
\caption{The parameter sizes and running time of GGNN, EGNN and NGNN on graphs with the number of nodes varying from 2 to 30. EGNN is a simple variant of GGNN, in which every edge has a different dtransformation matrix for propagating state information. }
\label{fig:complexity}
\end{figure}

\subsection{Model Analysis}

In this section, we analyze the parameter space and time complexity of our proposed method NGNN.

\subsubsection{Parameter Space}\label{sect:parameter}
The parameter space of our proposed method consists of two parts: the parameters correlated to nodes and the perception networks in attention mechanism.
For each node $n_{i}$, we have a linear mapping matrix $W_{h}^{i}$ to map the input features into the latent space, an input matrix $W_{in}^{i}$ and an output matrix $W_{out}^{i}$ to propagate state information. Totally we have $3\left | \mathcal{N}  \right |$ matrices, which are proportional to the number of nodes $\left | \mathcal{N}  \right |$. In addition, we have two matrices of perception networks in the attention mechanism. Overall, we have $3\left | \mathcal{N}  \right |$ matrices. Therefore, we say that the parameters of NGNN is node-wise. 

\subsubsection{Time Complexity}
For each input outfit $s$, the time complexity of training process in NGNN consists of three parts. The first part is calculating the transformed state information nodes receive, which is $O(\left | s \right |)$. The second part is updating the state of the concerned nodes, which is also $O(\left | s \right |)$.
The last part is updating the parameters of concerned nodes. Since the number of parameters correlated to the concerned nodes is $3\left | s \right|$ as described above, its complexity is also $O(\left | s \right |)$. According, the total complexity is $O(\left | s \right |)$.

We compare NGNN with GGNN \cite{li2015gated} and Edge-wise GNN (EGNN). EGNN is a straight-forward way to achieve edge-wise interactions. In EGNN, each edge has an unique transformation matrix for propagating state information. We train the three models on a toy dataset consisting of outfits with the number of items varying from 2 to 30. In another word, the three models are trained on graphs with the number of nodes varying from 2 to 30.
The parameter size and running time are shown in Figure \ref{fig:complexity}.
As can be seen, the parameter size and running time of EGNN increased quadratically while those of NGNN and GGNN increased linearly.

In summary, NGNN can better model edge-wise interactions and learn better node representations with appropriate parameter space and time complexity. When there are more training data, we can only update the parameters of concerned nodes offline and upload them online without changing the edges or other nodes, which is effective in real applications.

\section{Experiments}
As the main contribution of this work is to model outfit compatibility, we aim to answer the following research questions via experiments.
\begin{itemize}
\item[\textbf{RQ1}]
 How does our proposed NGNN perform compared with other pair or sequence representation models?
\item[\textbf{RQ2}]
 How do different components in NGNN effect the performance?
\item[\textbf{RQ3}]
 How do different modalities influence the inference of compatibility?
\item[\textbf{RQ4}]
 How do the hyper-parameters $d$ and $T$ effect the performance?
\end{itemize}
 Next, we first describe the two tasks we conduct experiments on and the experimental settings. We then report the results by answering the above research questions in turn.
 Finally, we analyze the parameter space and time complexity of our proposed model.

\subsection{Task Description}
We conduct experiments on two tasks: fill-in-the-blank fashion recommendation and compatibility prediction on the polyvore dataset described in Section \ref{sect:dataset}.

\subsubsection{Fill-in-the-blank Fashion Recommendation}
The task of fill-in-the-blank (FITB) in fashion recommendation is first introduced in \cite{han2017learning}. In this task,
Given a set of fashion items and a blank, we aim to find the most compatible item from the candidates set to fill in the blank as shown in Figure \ref{fig:fill_blank}. This is a practical scenario in people's daily life. For example, a person wants to select clothes to match the pants, hat and shoes.
Following \cite{han2017learning}, we create a fill-in-the-blank dataset based on the Polyvore test dataset. For each outfit in the test dataset, we randomly select an item from it and replace the item with a blank. By randomly select 3 negative items from $\mathcal{I}$ we can form a 4-length candidates set along with the positive one, assuming the negative items are less compatible than the positive one.
In our work, we respectively put the four candidate items into the blank to form an outfit and calculate their compatibility scores with our model. The one which achieves the highest outfit compatibility score is chosen as the answer.
The performance is evaluated by the accuracy of choosing the right answer from the four candidates (FITB accuracy), which is 25\% by randomly selecting.

\subsubsection{Compatibility Prediction}
This task is to predict the compatibility score for any given outfit.
To evaluate our model, we first construct a dataset $\mathcal{P}_{test}$ using the Polyvore test dataset.
Specifically, for each positive outfit $s$ in the Polyvore test dataset, we create a negative outfit $s^{-}$ by randomly selecting $\left | s \right |$ items from $\mathcal{V}$. Thus, we can obtain a dataset $\mathcal{P}_{test}$ consisting of evaluation pairs $\left ( s, s^{-} \right )$. We then adopt the widely used metric AUC (Area Under the ROC curve) to evaluate the performance, which is defined as,
\begin{equation}\label{eq1}
AUC = \frac{1}{\left | \mathcal{P}_{test} \right |}\sum_{\left ( s, s^{-} \right ) \in \mathcal{P}_{test}} \delta \left ( x_{s} > x_{s^{-}} \right ),
\end{equation}
where $\delta \left( a \right )$ is an indicate function
that returns one if the argument a is \emph{true} and zero otherwise.

\subsection{Experimental Settings}

\begin{table}[htbp]
  \centering
  \caption{The performance of models evaluated by Fill-in-the-blank accuracy and compatibility prediction AUC.}
    \begin{tabular}{ccc}
    \toprule
    Method & Accuracy (FITB)   &  AUC (Compatibility) \\
    \midrule
    Random &  24.07\%& 0.5104  \\
    SiameseCNNs & 48.09\%   & 0.7087 \\
    LMT   &  50.91\%  & 0.6782\\
    Bi-LSTM & 67.01\%  & 0.8427  \\
    GGNN (visual)& 72.78\%  &  0.9516 \\
    GGNN (textual)& 73.63\%  &  0.9591 \\
    \midrule
    NGNN (visual) & 77.01\%  & 0.9600 \\
    NGNN (textual) & 77.78\%  &  0.9716 \\
    NGNN (multi-modal)& \textbf{78.13\%}  & \textbf{0.9722} \\
    \bottomrule
    \end{tabular}%
  \label{tab:compared_method}%
\end{table}%

For optimization, we adopt the RMSProp optimizer. The optimal hyper-parameters (learning rate, batch size, trade-off parameter $\beta$, parameter for L2 regularization $\lambda$, hidden size $d$ and the propagation steps $T$) are determined by the grid search strategy. We experimentally find that the model achieves optimal performance with the learning rate as 0.001, batch size as 16, $\beta$ as 0.2, $\lambda$ as 0.001, $d$ as 12 and $T$ as 3.
We early stop the training process when the loss stabilizes, usually at 12 epoch. All the experiments were conducted over a sever equipped with 3 NVIDIA Titan X GPUs.

\subsection{Model Comparison (RQ1)}

To evaluate our proposed method, we compare it with the following state-of-the-art baselines. Note that SiameseCNNs and LMT focus on the compatibility between two items. In the task of outfit recommendation, we sum up the compatibility scores of all the pairs in the outfit as the score of the outfit.


\noindent\textbf{Random}: Randomly assign the compatibility scores of outfits.

\noindent\textbf{SiameseCNNs}: It's a pair-wise compatibility learning method \cite{veit2015learning}, which minimizes the distance of image vectors between compatible items and maximums the distance between incompatible items. This method only focuses on the visual information.

\noindent\textbf{LMT}: Proposed by \cite{mcauley2015image}, it models compatibility between items in a style space via a single low-rank Mahalanobis embedding matrix. This method also focuses on visual modality.

\noindent\textbf{Bi-LSTM}: It's proposed by \cite{han2017learning}. By viewing an outfit as a sequence, it uses the bidirectional LSTMs to predict the next item, taking multi-modal data as input, which can also predict the compatibility scores of outfits. This method focus on multi-modal information.

 \noindent\textbf{GGNN (visual / textual)}: We use GGNN to model the outfit compatibility with input of visual or textual features of items.

\noindent\textbf{NGNN (visual / textual)}: We use our proposed NGNN to model the outfit compatibility with input of visual or textual features of items.

\noindent\textbf{NGNN (multi-modal)}: We jointly models the outfit compatibility with NGNN from the visual and textual modalities.

\begin{figure}[t]
  \centering
  \includegraphics[scale=0.75]{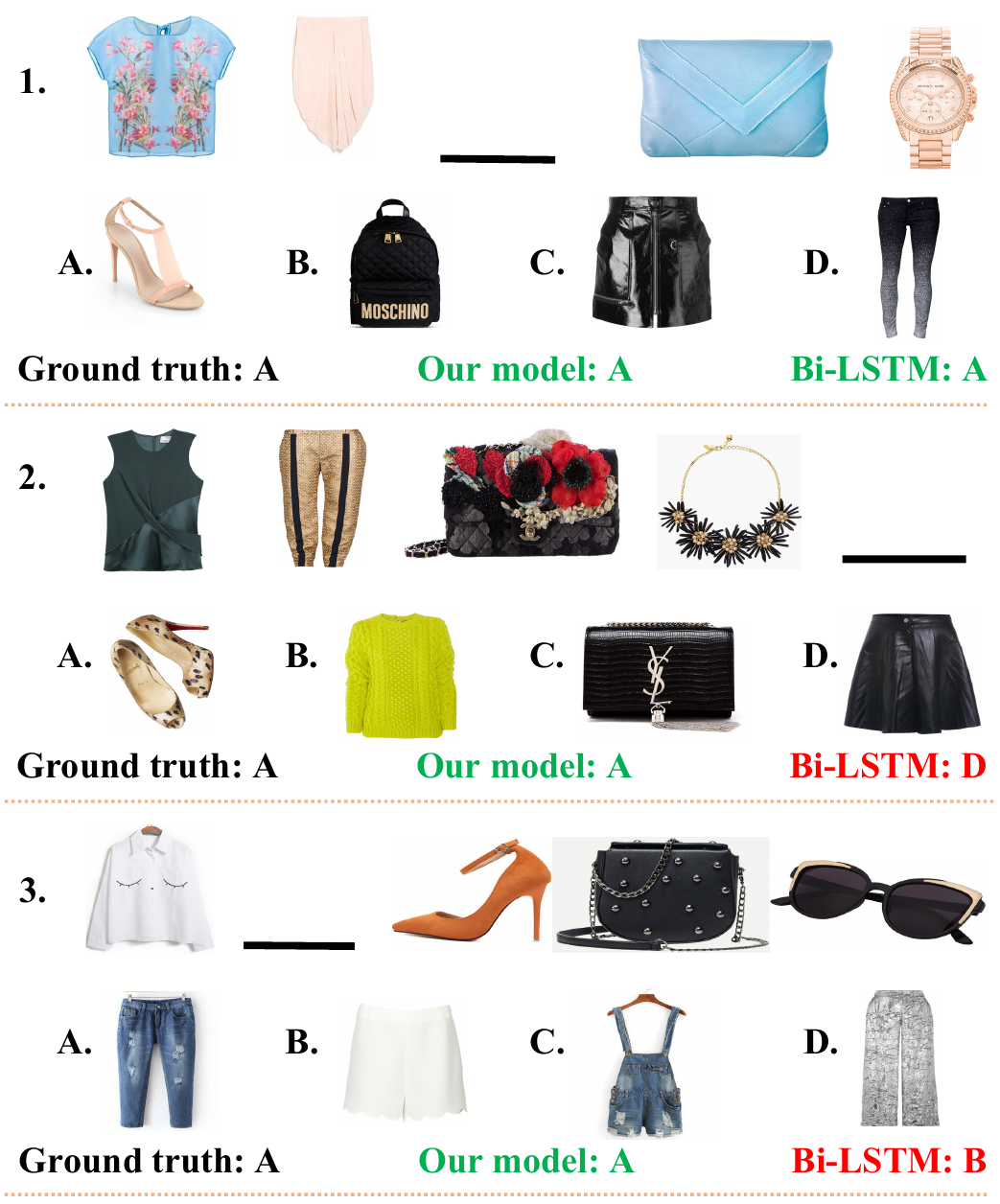}
  \caption{Comparison of our model and Bi-LSTM on several examples in fill-in-the-blank task. Green fonts represent the right chosen and red fonts represent wrong chosen. }\label{fig:fill_blank}
\end{figure}

We conduct comparison experiments on the tasks of fill-in-the-blank fashion recommendation and compatibility prediction. The results are shown in Table \ref{tab:compared_method}. Next we analyze the results of FITB accuracy and compatibility AUC respectively.

\subsubsection {FITB Accuracy} \label{sect:fitb}
The comparison of FITB accuracy is shown in the middle column of Table \ref{tab:compared_method}. We can obtain the following observations from the table:
(1) SiameseCNNs and LMT achieve worse performance compared with the other methods, which demonstrates that the pair representation ignores the integrity of outfit. (2) GGNN outperforms Bi-LSTM, suggesting that our proposed graph representation can better reflect the complex relations among items in an outfit than sequence representation. (3) NGNN achieves a better performance than GGNN, which proves that our proposed NGNN can better infer compatibility information from the graph by better modeling node interactions.
(4) Using textual features achieves better performance than visual features for both GGNN and NGNN, which suggests that the textual features can better represent fashion items than visual features in that the titles accurately summarize the key features of items.
(5) NGNN with Multi-modal input achieves best performance, proving that the visual and textual information are good supplements to each other. In addition, our proposed method can jointly model the outfit compatibility from multiple modalities.
\begin{figure}[t]
  \centering
  \includegraphics[scale=0.40]{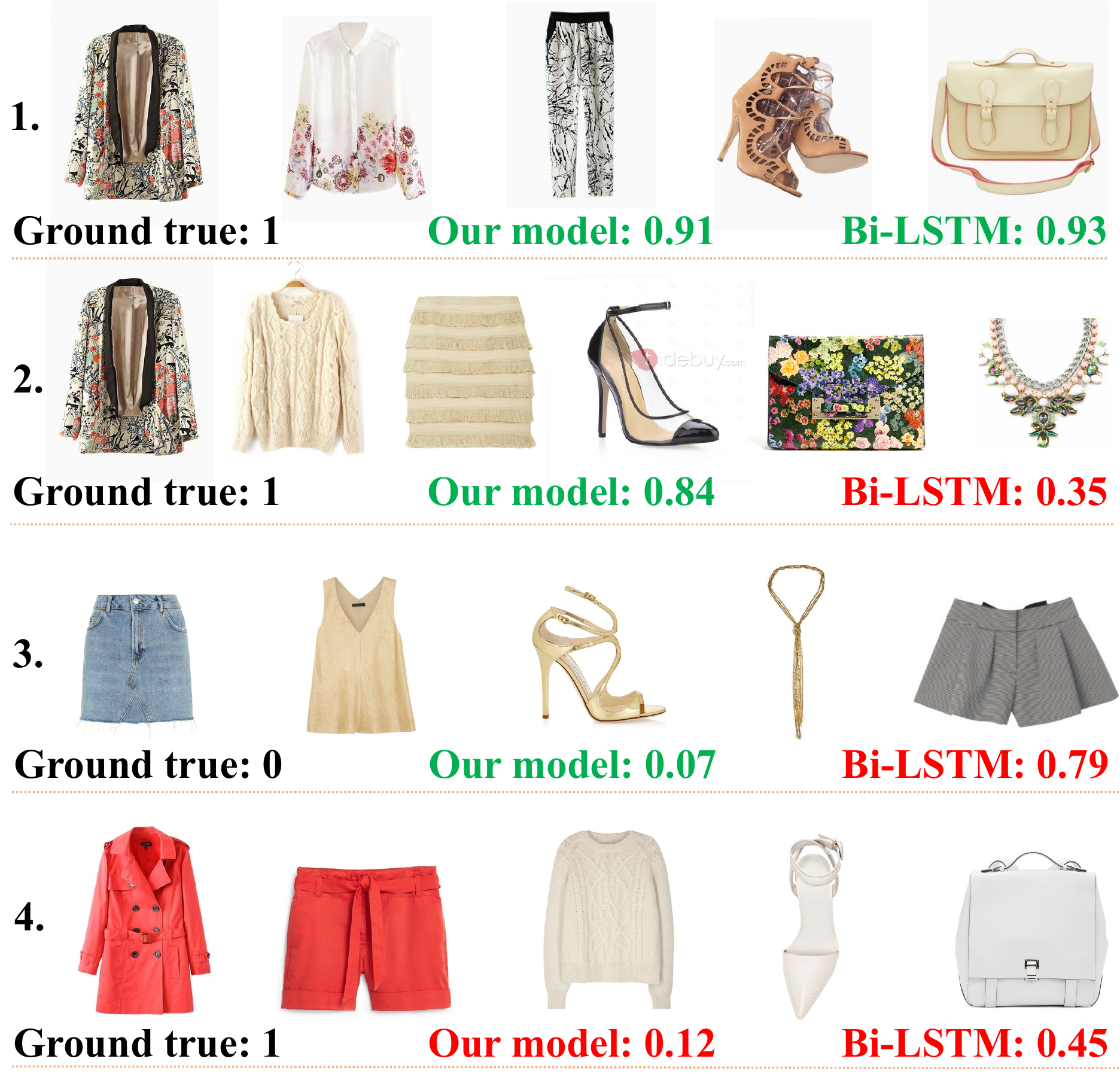}
  \caption{Comparison of our model and Bi-LSTM on several example of task compatibility prediction. Green fonts represent the right evaluation and red fonts represent wrong.}
  \label{fig:case_study}
\end{figure}
We randomly select several example outfits in the FITB dataset to test on our model and Bi-LSTM. The results are shown in Figure \ref{fig:fill_blank}.
From example 1, we can see that Bi-LSTM and our model can both correctly select the item of complementary category as the outfit lacks of a pair of shoes. In example 2, our model chooses the right item while Bi-LSTM choose the skirt wrongly, which conflicts with the pants. This may be because that the position of pants is far away from the blank so that it has little influence on it. In example 3, our model selects the most compatible bottom jeans with the whole set from the four bottoms while Bi-LSTM selects the white short skirt which may be most compatible with the white shirt, but not such compatible with the other items. The wrong chosen of Bi-LSTM may be because that the nearby white shirt has strong influence on the blank.
From these example, we can draw a conclusion: the items have strong influence on nearby items in the sequence while have less influence on items faraway. This makes the sequence represented method relies on a proper sequence order.
	Compared with the sequence-represented model Bi-LSTM, our graph-represented model can better model the complex relations among items in an outfit.

\subsubsection{Compatibility AUC}
The performance comparison of models evaluated by compatibility AUC is shown in the third column of Table \ref{tab:compared_method}.
Similar with the comparison of performance evaluated by FITB accuracy, NGNN achieves best performance, GGNN is second best, and then is the sequence represented method Bi-LSTM, the pair represented methods (SiameseCNNs, LMT) are worst. Using textual features can achieve better performance than visual features. NGNN with multi-modal data achieves the best performance. These observations support our analysis in Section \ref{sect:fitb}.
In addition, it can be seen that GGNN and NGNN all achieve great performance and the gap is small, which proves the superiority of graph representation on compatibility prediction.
GGNN is already good enough for compatibility prediction.

We randomly select several example outfits and show the compatibility scores predicted of Bi-LSTM and our model in Figure \ref{fig:case_study}.
In example 1, our model gives the outfit a high score correctly while Bi-LSTM gives it a relatively low score, which may be because that the sequence representation method fails to capture the high compatibility between the first top and the fifth handbag.
While Bi-LSTM and our model both give the outfit a high score correctly in example 2.
In example 3, the fifth shorts conflict with the first skirts. Our model gives it a low score correctly while Bi-LSTM gives it a relatively high score, which may be because Bi-LSTM fails to capture the conflict of the two bottoms across a long distance. Examples 2 and 3 suggest that Bi-LSTM relies on a proper order of sequence, which is flexible and hard to learn. The influence between two items is inversely proportional to their distance in the sequence. Our model can better reflect the relation among items compared with Bi-LSTM.
In example 4, our model gives it a low score to the positive outfit. A plausible explanation is that there is little matching of coat, sweater and shorts in training dataset.

\begin{table}[htbp]
  \centering
  \caption{The performance of our proposed method with different components.}
    \begin{tabular}{ccc}
    \toprule
    Method & FITB Accuracy  &  Compatibility AUC \\
    \midrule
    NGNN(-W-A)& 75.17\%   & 0.9541 \\
    NGNN(-W)& 75.71\%  & 0.9559\\
    NGNN(-A)& 76.12\%   & 0.9578 \\
    NGNN& \textbf{77.01}\%  & \textbf{0.9600} \\
    \bottomrule
    \end{tabular}%
  \label{tab:ablation}%
\end{table}%

\subsection{Component Comparison (RQ2)}

To investigate the effect of different components on the performance of our proposed method, we compare our method with different components.

\noindent\textbf{NGNN (-W)}:
NGNN without weighted adjacency matrix. Specifically, we use 0-1 adjacency matrix instead of the weighted one as described in Section \ref{sect:gi}.

\noindent\textbf{NGNN (-A)}: NGNN without attention mechanism. We only use a perception network to model the compatibility scores of nodes and sum them up them.

\noindent\textbf{NGNN (-W-A)}: NGNN with neither the attention mechanism nor weighted adjacency matrix.

The performances are shown in Table \ref{tab:ablation}. If not specifically stated, we use visual features as input of NGNN. We observe that NGNN outperforms all the ablative methods, which proves the necessity of the two components in our model.
NGNN (-W) achieves worse performance than NGNN, suggesting that the weighted adjacency matrix we calculated is a supplement to the information of graph. NGNN outperforms NGNN (-A), which confirms the necessity of our attention mechanism.

\begin{figure}[htbp]

  \centering
  \includegraphics[scale=0.40]{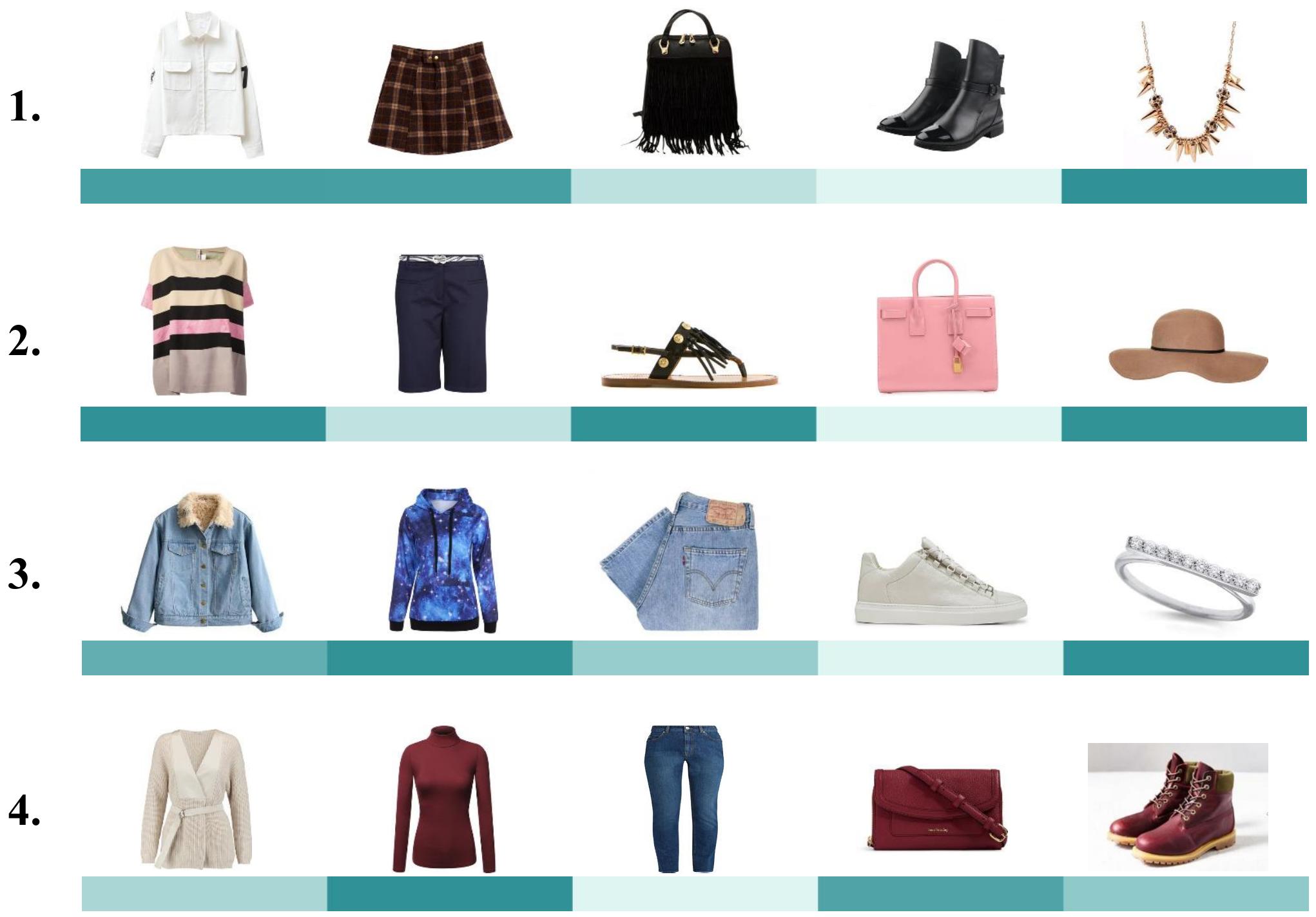}

  \caption{Illustration of attention mechanism. The depth of color are proportional to the attention weights of items.}\label{fig:atten}

\end{figure}

To intuitively illustrate how the attention mechanism works, we show the attention weights of several example outfits in Figure \ref{fig:atten}. The depths of color are proportional to the attention weight of items. The higher attention weight item has, the deeper the color of bar below it is, the stronger influence the item has to the outfit compatibility.
As can be seen in example 1, the shirt, short skirt and necklace have distinct and compatible styles and they have strong influence to the outfit compatibility. The black bag and black shoes have little influence to outfit compatibility while they have ordinary styles. Similarly, the sweater, shoes and hat in example 2 are all of distinct rustic style and they have strong influence to the outfit compatibility. The jeans and shoes in example 3 can match well with most styles while they have little influence to the outfit compatibility. The white outwear and jeans which can match with many styles in example 4 also have little influence to outfit compatibility.
From these examples, we can find that more distinct style items have, more influential they are to the outfit compatibility. On the contrary, more ordinary items are (i.e., can match with many styles), less influential they are. This is reasonable in common sense. 

\subsection{Modality Comparison (RQ3)}

 \begin{figure*}[t]
  \centering
  \includegraphics[scale=0.58]{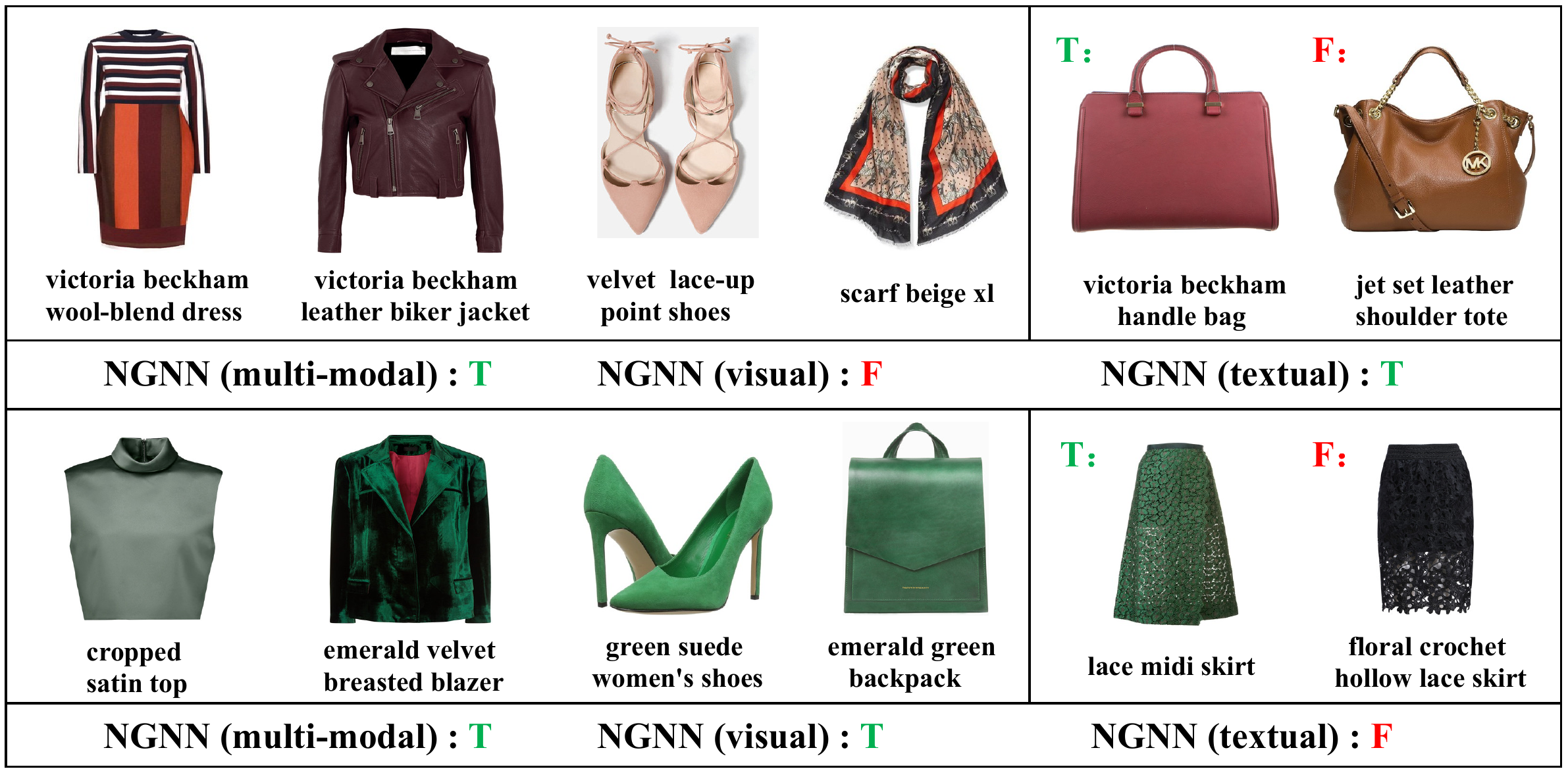}
  \caption{Comparison of NGNN(visual), NGNN(textual) and NGNN(multi-modal). For each outfit in the left, there is a true answer (T) and a false answer (F). Green fonts reprsent true chosen while red fonts represent false chosen.}
  \label{fig:modal_compare}
\end{figure*}

\begin{figure}[hbtp]
  \centering
  \vspace{1.0em}
  \includegraphics[scale=0.68]{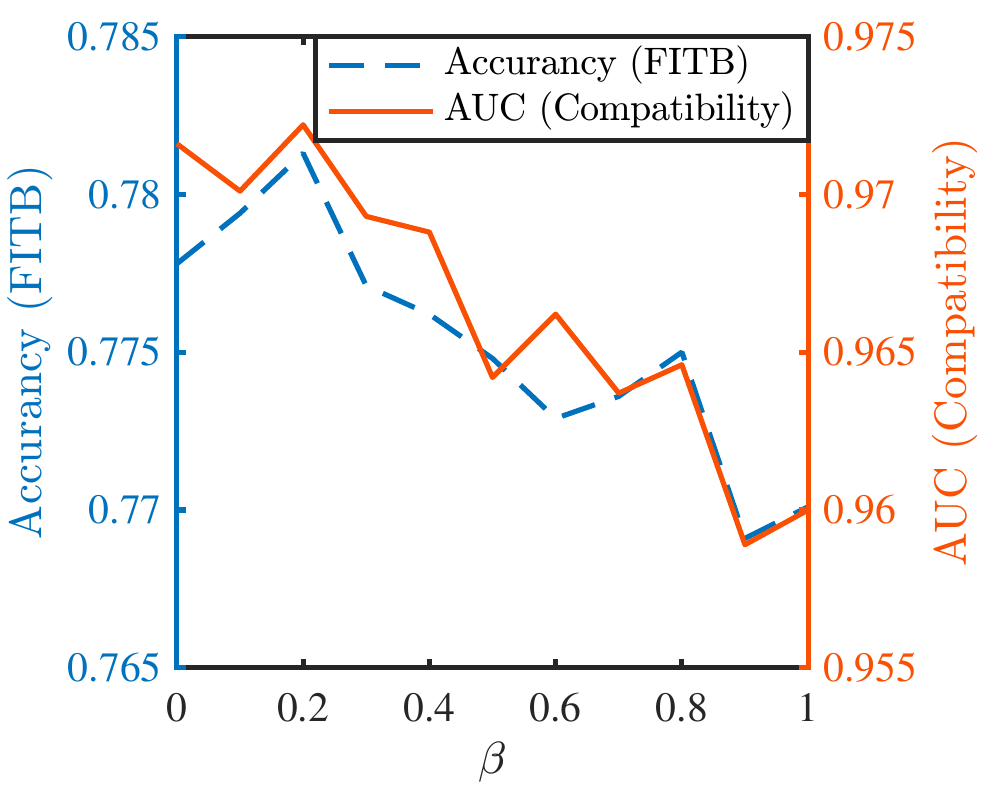}

  \caption{FITB accuracy and compatibility AUC of our model with different $\beta$.}
  \vspace{1.0em}
  \label{fig:beta}
\end{figure}

\begin{figure}[hbtp]
\centering

\subfigure[State dimensionality]{
\begin{minipage}[b]{0.24\textwidth}
\label{fig:cloth1} 
\includegraphics[width=1\textwidth]{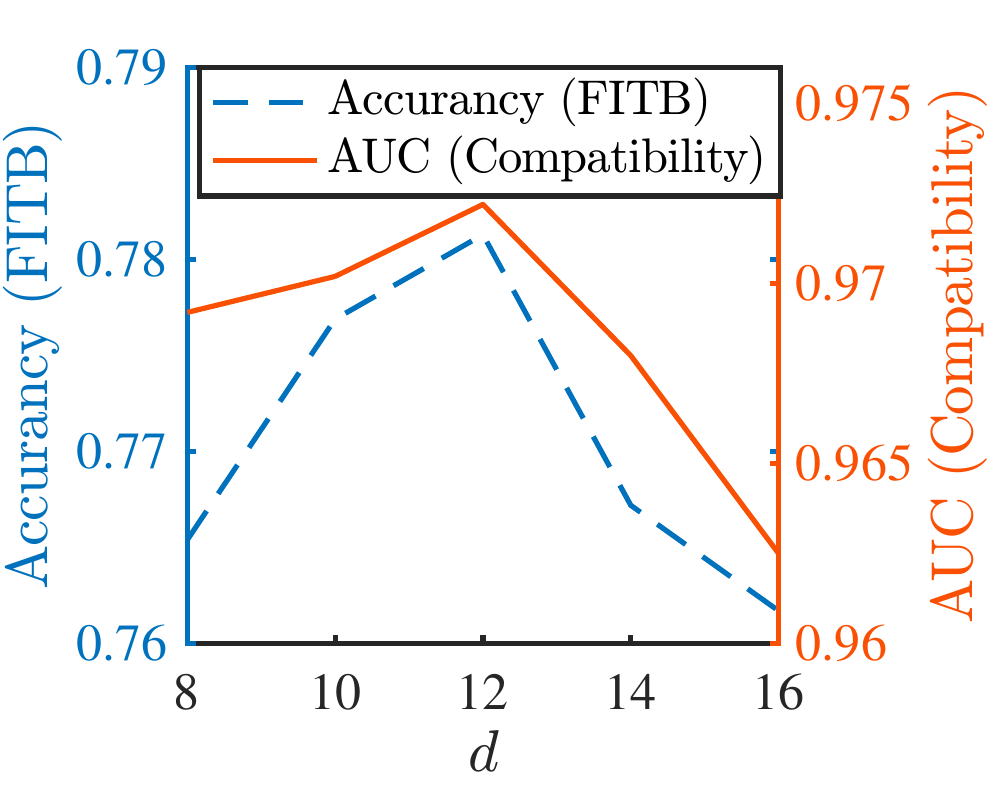}

\end{minipage}%
}%
\subfigure[Propagation times]{
\begin{minipage}[b]{0.24\textwidth}
\label{fig:cloth2} 
\includegraphics[width=1\textwidth]{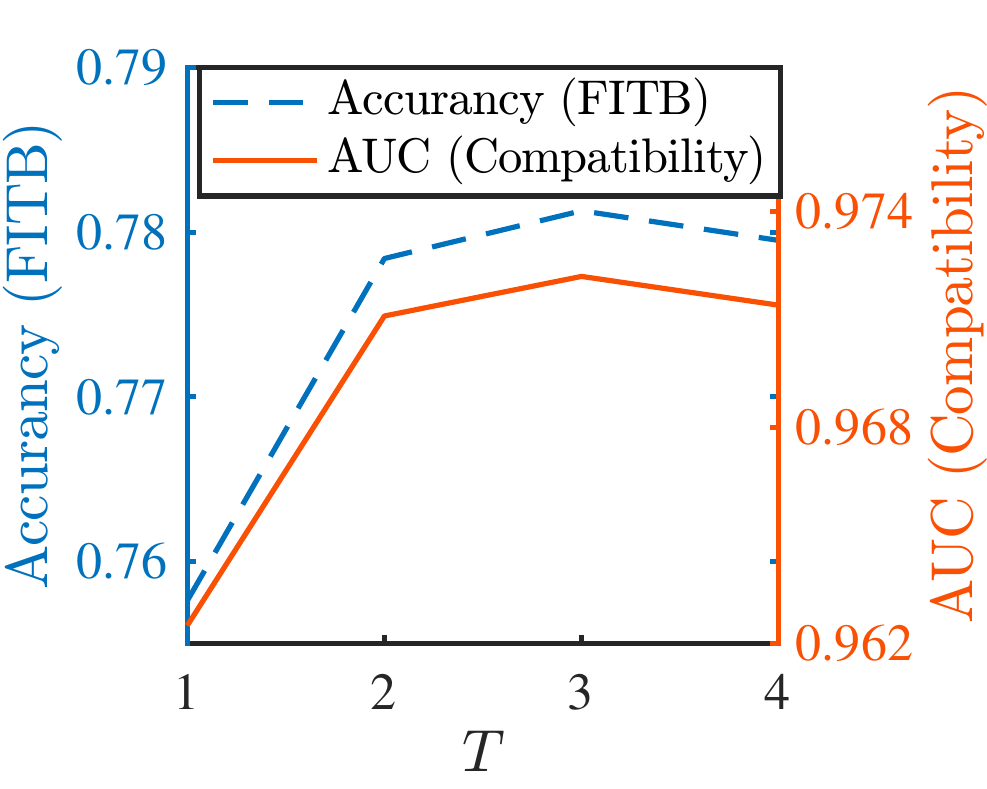}
\end{minipage}%
}%
\caption{AUC and FITB accuracy performance with different dimensionality $d$ (left) and propagation times $T$ (right).}
\label{fig:performance}
\end{figure}

To investigate the influence of different modalities on the performance, we compare the NGNNs (multi-modal) with different modality combinations.
As described in Equation \ref{compatibility}, $\beta$ is the weight of visual compatibility score while that of textual compatibility score is $1-\beta$. Note that when $\beta=0$ the model is actually NGNN (visual), when $\beta=1$ it's NGNN (textual).
In Figure \ref{fig:beta}, we show the performance of NGNN (multi-modal) with $\beta$ varying from 0 to 1.
It can be seen when $\beta=0.2$, our model achieves the best performance, which suggests that the visual and textual information are good supplement to each other and both contribute s to the inference of compatibility. We can also see that using textual features can achieve better performance than visual features. A plausible explanation is that the textual information is more concise to capture the key features of items.

To intuitively see the difference of NGNN with visual, textual and multi-modal input, we show the comparison of them on several testing examples in Figure \ref{fig:modal_compare}.
In the example 1, Visual NGNN chose the false answer due to its similar color and material with the jacket, which the visual information can reflect. However the similarity between the textual information (titles) of the true answer and the dress, jacket directly reveals their high compatibility.
In example 2, it's hard to choose the answer with textual input since the titles of the two answers are similar, while with visual input  NGNN can distinct the high compatibility between the styles of the true answer and the outfit from visual information.
In summary, the visual and textual information are great supplements to each other. By integrating them we can achieve better performance.

\subsection{Hyper-parameter Discussion (RQ4)}

In this section, we examine how the dimensionality of hidden vectors $d$ and the propagation time $d$ affect the performance. Comparison experiments are conducted on NGNN (multi-modal).
Figure \ref{fig:performance} shows the performance of our model with varying $d$ and $d$.
From the figure in the left, we can observe that the FITB accuracy and compatibility AUC reach the highest point when $d$ is around 12. When $d$ is large than 12, the performance drops sharply. This indicates that there is no need to represent the node state with too much parameters.
From the figure in the right, it can be seen that the performance saturates when $T$ is $3$ and the nodes in NGNN have interacted with others sufficiently, suggesting that it's important to model the multi-hop propagation so that the nodes are better aware of other nodes' information. Moreover, when $T$ is high, the superfluous information propagation will confuse the nodes.


\section{Conclusion}
In this paper, we first propose to model the outfit compatibility by representing an outfit as a graph since the graph structure can better capture the complex relations among items in outfits.
To infer the compatibility information from the graph, we further propose a novel model NGNN, which can better model flexible node interactions and learn better node representations.
Our proposed method can model outfit compatibility not only from visual or textual modality respectively, but also from multiple modalities jointly.
Experimental results on a real-world dataset (Polyvore dataset) prove the great superiority of our method over existing works.
In the future, we aim to model personalized compatibility since people have different notions of fashion compatibility.

\begin{acks}
This work is jointly supported by National Natural Science Foundation of China (61772528, 61871378) and National Key Research and Development Program (2016YFB1001000).
\end{acks}

\bibliographystyle{ACM-Reference-Format}
\balance
\bibliography{sample-iw3c2}

\end{document}